\documentclass[aps,pre,preprint,groupedaddress]{revtex4}
\usepackage{graphicx,graphics} 
 
\begin{document} 
 
\title{Conservation of statistical results under the reduction of pair-contact interactions to solvation interactions} 
 
\author{N. Hamedani Radja\footnote{To whom correspondence may be addressed. E-mail: nima@mehr.sharif.edu or ejtehadi@sharif.edu}$^\dag$ } 
\author{R. R. Farzami$^\dag$} 
\author{M. R. Ejtehadi$^{* \dag \ddag}$} 
 
\affiliation{$^\dag$Department of Physics, Sharif University of 
Technology, P.O.Box 11365-9161, Tehran ,Iran. \\ 
$^\ddag$Institute for studies in Theoretical Physics and 
Mathematics (IPM), P.O.Box 19395-5531, Tehran, Iran. } 
 
\begin{abstract} 
 
We show that the hydrophobicity of sequences is the leading term 
in Miyazawa-Jernigan interactions. Being the source of additive 
(solvation) terms in pair-contact interactions, they were used to 
reduce the energy parameters while resulting in a clear vector 
manipulation of energy. The reduced (additive) potential performs 
considerably successful in predicting the statistical properties 
of arbitrary structures. The evaluated designabilities of the 
structures by both models are highly correlated. Suggesting 
geometrically non-degenerate vectors (structures) as protein-like 
structures, the additive model is a powerful tool for protein 
design. Moreover, a crossing point in the log-linear diagram of 
designability-ranking shows that about $\frac{1}{e}$ of the 
structures have designabilities above the average, independent on 
the used model. 
 
\end{abstract} 
 
\maketitle 
 
\section{Introduction} 
 
The challenge to understand the different aspects of the protein 
folding phenomenon has been brought further by means of different 
disciplines, specially after the classic contribution by 
Anfinsen~\cite{56}. The extreme complexity of the problem due to 
the numerous number of interacting atoms has made the study of 
the problem practically impossible in the atomic scales. The 
simplified coarse-grained models introduce a feasible way to 
understand significant macroscopic characteristics of the 
macromolecules without any reference to the atomic details. 
However, a relatively impeccable estimation of the effective 
inter-residue interactions is necessary for such models in order 
to yield reasonable results. 
 
Although it is believed that the solubility of different residues 
in the water environment (that is, a single-body effective 
solvation potential) plays the most significant role in the 
folding process, the most commonly used class of potentials in the 
folding problem is the class of two-body interactions e.g. the 
potential introduced by Miyazawa and Jernigan (MJ)~\cite{93}. 
Introducing an effective contact distance and paying particular 
attention to the contacts between different residues in a 
protein's native structure, they reached an estimation of the 
effective inter-residue interactions by means of a statistical 
method~\cite{93}. The contact energy parameters $M_{ij}$ form a 
$20\times20$ symmetric matrix, namely the 
MJ~matrix~($\mathbf{M}$), presenting $210$ independent contact 
energy parameters between $20$ known amino acids. This matrix has 
been modified later by considering a larger database of 
proteins~\cite{52} and through some self consistent iterative 
procedures~\cite{157,113,201}. Although the MJ parameters are not 
particularly successful in the study of the folding dynamics and 
mechanism, they predict the groundstate configuration from decoys 
quite fine~\cite{201}. 
 
Analyzing the eigenvalues of MJ matrix, Li {\it et al.} have shown 
that its elements can be approximated quite fine by considering 
only $22$ parameters instead of the original $210$ 
parameters~\cite {21}. That is $20$ energy parameters, $q$, one 
residue independent mixing energy parameter, $\gamma$, and a 
collapsing energy parameter, $E_c$, which is physically 
irrelevant in the energy spectrum of the structures sharing the 
same compactness. The pair-contact energy between two residues 
having energy parameter $q$ and $q^{\prime}$ is given by: 
\begin{equation} 
\label{gamma} E_{q q^{\prime}} = -(E_c + q + q^{\prime} + \gamma q 
q^{\prime}) 
\end{equation} 
in their model. They argued that the energy parameters ($q$'s) are 
measures of hydrophobicity of residues and their distribution 
supports simplified two-letters Hydrophobic-Polar (HP) 
model~\cite{124}. The authors~\cite{21} suggested the HP energy 
parameters as $1$, $0$ and $0.3$ for $q_H, q_P$ and $\gamma$ 
respectively, in an arbitrary energy unit. Furthermore, they have 
shown that there always exist a set of highly designable 
structures which are reasonably better candidates as groundstate 
structures compared to the other configurations, regardless of 
using the energy parameters either from the HP model~\cite{22} or 
the 20-letters model (MJ energy parameters)~\cite{11}. 
 
In a recent work~\cite{250} it has been shown that most commonly 
used pair-contact potentials can be divided to two different 
class of small and large $\gamma$ values. Setting $\gamma=0$ in 
Eq. (\ref{gamma}), we are left with the additive part of the 
contact potential energy which may be treated as a solvation-like 
model~\cite{92,120}. In this reduced model, the pair-contact 
energy of a given protein with $N$ residues in a given 
configuration may be represented using vectorial notation in a 
compact fashion: 
\begin{eqnarray} 
E(\{ q_i\},\{\vec{r}_{i}\} ) &=& \frac{-1}{2} \sum_{i,j}^{N}{(q_i+ 
q_j) 
C_{ij}} \nonumber \\ 
&=& - \sum_{i,j}^{N}{q_i C_{ij}} \\ 
&=&  -\vec{q}\cdot \vec{v}, \nonumber 
\end{eqnarray} 
where $i$ and $j$ are residue's indexes along the chain, 
$\vec{r}_{i}$ is the position of residue $i$ and $\mathbf{C}$ is 
the structure's contact matrix. The element $C_{ij}$ is $1$ if 
the distance between the corresponding non-chemically bonded pair 
in the given structure is smaller than a cut-off radius, $r_c$, 
and is zero otherwise. The vector $\vec{v}$, defined as 
$v_i=\sum_{j=1}^{N}{C_{ij}}$, is a one dimensional projection of 
the contact matrix and is called the contact vector~\cite{92,46}. 
The components of this integer vector are simply the number of 
contacting neighbors for corresponding residues, regardless to 
their types. Although the reduced model decreases the 
computational cost as it reduces matrix products to vector 
products, this presentation carries noticeably less information 
compared to the contact matrix representation. The vectors are 
generally degenerate and any of them may correspond to $g$ 
different contact matrices. The case in which $g=1$, the vector 
is non-degenerate and there is a one to one relationship with its 
corresponding matrix. Recent studies show that the 
three-dimensional structure is reproducible using one-dimensional 
vector representation for some small proteins~\cite{29,163} which 
suggest that the contact vectors of protein-like structures have 
to be non-degenerate. 
 
Consistent with many arguments about the central role of 
hydrophobic forces in the protein folding 
problem~\cite{94,95,96}, it has been shown that in the case of 
two-letters (HP) lattice model, the additive (solvation) terms of 
the potential carry enough information to replicate the energy 
spectrum of sequences in the configuration space~\cite{92}. It 
has also been shown that non-degenerate contact vectors ($g=1$) 
are noticed in parallel to highly designable structures as well 
as large energy~gaps between the groundstates and the first 
excited states of native proteins~\cite{92,4}. 
 
In this study, we show that in the case of 20-letters model, the 
additive part of MJ matrix carries enough information to manifest 
highly designable structures, by enumerating all chain 
configurations in a two dimensional square lattice model up to a 
length of 36. Although the applicability of the additive 
potential instead of MJ potential is not very promising in coming 
up with the same groundstate structure of a randomly given 
sequence, its performance is appealing on protein-like sequences. 
 
\section{The Model} 
 
We consider the MJ96 interaction matrix ($\textbf{M}$) 
(introduced by Miyazawa and Jernigan at 1996~\cite{52}) as the 
reference pair-contact interaction matrix between all different 
types of the residues. The additive potential defined as: 
\begin{equation} 
\label{add} E_{ij}=q_{i}+q_{j} 
\end{equation} 
will be regarded as the first-order approximation of the 
interaction matrix in which indices $i$ and $j$ (going from $1$ to 
$20$) refer to the residue types. Turning back to the 
approximation made in~\cite{21}, one can identify the additive 
potential as the $\gamma=0$ limit of Eq. (\ref{gamma}) which may 
still hold the general behavior of the original model. A similar 
procedure has been done in the case of the two-letters HP 
model~\cite{92}. The other constant parameter, $E_{C}$, is 
absorbed into the residue dependent parameters. 
 
The root mean square distance of the two matrices, defined as 
${\rm RMSD}=\sqrt{\sum_{ij}{(M_{ij}-E_{ij})^{2}}}$ has been taken 
as a measure of the similarity between them. Therefore, a set of 
additive parameters ($q$'s) minimizing the RMSD of the two 
matrices will theoretically mimic the original inter-residue 
interactions (MJ matrix). Minimizing RMSD, simply gives $q$'s as: 
\begin{equation} q_{i}=\langle M 
\rangle_{i}-\frac{1}{2}\langle M \rangle, 
\end{equation} 
where the first term is the average of the strength of 
interactions between $i$th amino acid and all other ones 
(including itself) while the second term is a residue independent 
value, equal to the half of the mean of interactions between all 
different residue types which may be described as a tendency for 
proteins to fold into compact structures. For the structures with 
the same compactness (having the same number of contacts), it 
causes an irrelevant shift in energy spectrum. 
 
To find the sensitivity of low-energy states to the model 
interaction energies, we enumerated all maximally compact 
structures having different lengths of $L=16, 20, 30, 36$. We 
also studied the whole structures with $L=14, 16$ to be sure that 
the compactness restriction does not affect the conclusions. 
Working in complete configuration space of larger chains is not 
feasible in a reasonable time. The time limitation also 
restricted us to consider only two-dimensional structures. Among 
all structures which are identical by geometrical symmetries, we 
chose only one. 
 
Because of the large number of possible sequences, we restricted 
our evaluation to an ensemble of random sequences, followed by 
search for their groundstates and first excited states in the 
structures space. To be more specific, in the case of $L=36$ we 
sampled $2.4 \times 10^7$ different sequences among 
$20^{36}\simeq10^{47}$ possible sequences and $10^7$ for other 
cases. This amount of the samples yield to valid statistical 
results as there was no significant impact on the results even 
after ignoring half of these samples. 
 
Starting with the MJ96 model as the reference potential, we 
searched the conformation space for the groundstate and the first 
excited state of each of the sampled sequences using the 
pair-contact interactions. Moreover, we evaluated the energy of 
these two structures using the additive potential. A similar 
search in the conformation space for the groundstate and the first 
excited state of each sequence is followed using the additive 
potential. We did the same comparison with the MJ85 
pair-interactions~\cite{93} and also for its additive part. 
 
\section{Results} 
 
\subsection{The D matrix} 
 
It is well believed that the hydrophobicity of the residues plays 
a very important role in the folding process. However, there is 
no unique scale for measuring this characteristic. Although the 
different scales agree with each other, they are strongly 
sensitive to the type of the experiment (Table~\ref{reg}). As it 
was mentioned before, the additive potential parameters ($q$'s) 
extract the contribution of hydrophobicity in MJ matrix. 
Therefore, they may be regarded as a new measure of 
hydrophobicity. We have compared them with the experimental 
scales and also with the q parameters, found by Li \textit{et 
al}~\cite{21} in Table~\ref{reg}. The upper half of the table 
represents the correlation coefficient ($r$) between 
hydrophobicity scales where the lower half is dedicated to the 
corresponding p-values (a measure of significance of the 
statistics). The table shows a significant correlation between 
the additive part of MJ matrix and the other hydrophobicity 
scales. This correlation has been observed formerly by 
Chan~\cite{121} in a reverse way, i.e he noticed a correlation 
between the elements of MJ matrix and the sum of the 
hydrophobicity of the corresponding amino acids. 
 
The reconstructed interaction matrix, Eq.(\ref{add}), is quite 
similar to the original MJ matrix. The elements of these two 
matrices have been compared with each other in 
Figure~\ref{mjaddmat}, showing a strong correlation of $r=0.982$ 
which is even stronger than the correlation between the different 
revisions of the MJ matrix ($0.973$) (introduced in 
1985~\cite{93} and 1996~\cite{52}). It means that the difference 
between the additive matrix and the reference one is in the range 
of statistical errors. Interestingly, the additive parts of the 
two MJ matrices correlate better than the matrices themselves 
($r=0.979$ in this case). 
 
Figure~\ref{gray} is a graphical representation of the matrices 
using a gray-scale intensity plot, i.e darker elements correspond 
to stronger interactions. As one can conclude, the matrices are 
quite similar to each other and the relative difference of each 
element is small, compared to the element itself. Roughly 
speaking, the left graph (additive) is the faded version of the 
right one (MJ) in Figure~\ref{gray}. Therefore, we can consider 
the differences as perturbative terms to the additive matrix. 
This result agrees with the conclusion of~\cite{92} where 
parameter $\gamma$ was introduced as a perturbation to the HP 
model. 
 
We define the difference matrix as: 
\begin{equation} 
D_{ij}=M_{ij}-( q_{i}+q_{j} ). 
\end{equation} 
 
We have sketched the distribution of the elements of the matrix D 
in Figure~\ref{diff} in order to see how large they are. Except a 
few large elements, the difference of the two matrices is small, 
compared to $k_{\rm B}T$. The energy of a sequence in a specific 
structure consists of the interaction energy of all of its 
contacts, meaning that this cumulative error may be not as 
negligible as of a single contact. Based on the Central Limit 
Theorem, the mean and the standard deviation of the sum of $N$ 
randomly chosen elements of $\mathbf{D}$ are larger than mean and 
the standard deviation of its elements, by factors $N$ and 
$\sqrt{N}$ respectively. For example, highly compact $2D$ 
structures with $36$ residues on a square lattice have $25$ 
contacts between the residues. The above discussion suggests that 
employing the additive matrix instead of the MJ matrix is 
plausible, together with a random energy with a zero mean and a 
standard deviation of $1.40k_{\rm B}T$. This random deviation is 
large enough to make the groundstates unstable. However, we will 
show that the chain correlations and the correlated shifts in the 
energy spectra considerably moderate the above na\"{\i}ve 
estimation. 
 
\subsection{The statistical properties of structures 
with different lengths} 
 
As mentioned before, the vector representation of configurations 
is a tricky method for decreasing the CPU run time as it reduces 
the matrix calculations to vector calculations. It is also 
responsible for some new degeneracies. In other words, in the 
vector model, all of the structures belonging to a vector with 
$g>1$ have the same energy. Thus, none of them can correspond to 
an unique groundstate. Table~\ref{Lattice} reports the number of 
contact matrices, contact vectors, structures with unique contact 
vectors and also the average of $g$ for different lengths, and 
Figure~\ref{lenghts} shows the first three of these parameters in 
a log-log plot. Although the number of vectors with $g=1$ is 
increasing with length, its ratio to the total number of 
structures is decreasing. Thus, the average value of $g$ is an 
increasing function of length, implying that the probability of 
being a protein-like structure is smaller than that of short 
lengths for larger structures. 
 
\subsection{Protein-like sequences} 
 
Searching for the groundstate of any given sequence using the MJ 
energy parameters, we will deal with one of the following 
scenarios: 
 
\begin{enumerate} 
 
\item 
 
\textit{It does not have a global minimum energy configuration:} 
It is not a protein-like sequence and is not a matter of our 
interest. 
 
\item 
 
\textit{It has a global minimum energy configuration but the 
structure corresponds to a vector with $g>1$:} There are $g-1$ 
other configurations which have the same form in vector 
representation. Thus this sequence does not have a unique 
groundstate using the vector (additive) model. It has been shown 
that in the case of HP lattice model, such sequences are of 
little interest and do not behave as protein-like 
sequences~\cite{92}. More careful considerations are required to 
conclude the same statement for 20-letters models. 
 
\item 
 
\textit{It has a global minimum energy configuration and the 
structure corresponds to a vector with $g=1$:} In the HP model, 
such structures are usually highly designable. Folding in highly 
designable structures, such sequences will be called protein-like 
sequences. 
 
\end{enumerate} 
 
In order to find the frequency of occurrence of each of the above 
scenarios, we have studied a set of randomly chosen sequences. 
Table~\ref{scenarios} shows the ratio of sequences belonging to 
each scenario. The average energy gaps, the energy differences 
between the groundstates and the first excited states, are also 
reported for different scenarios. The sequences belonging to the 
third scenario have a considerably larger energy gap which is yet 
another observation relating the protein-like structures to those 
having non-degenerate vector representations ($g=1$). The 
probability of occurrence of the first scenario remains almost 
constant for different lengths but the probability of belonging 
to the second and third scenarios increases and decreases with 
their lengths, respectively. It means that fewer sequences have 
the chance to fold in a structure with $g=1$, i.e. those which 
are claimed to be the protein-like sequences. 
 
\subsection{A comparison between the observable parameters of the 
reference and the additive potential} 
 
The aim of this section is to use the additive potential as an 
approximation to the reference pair-contact potential. Thus, a 
relevant question is how successful it is in finding the 
groundstate structures. Our investigation shows that using the 
MJ96 potential as the reference and for the structures with length 
of $36$, about $33\%$ of sequences belonging to the third 
scenario fold in the same structures, once we employ the additive 
approximation. Apparently, it seems that the additive 
approximation is not promising in the prediction of the 
groundstates. However, comparing the two versions of the MJ 
potential (MJ96 and MJ85), the overlap between the groundstates 
of these sequences is found to be only 43\%. 
 
It should be noted that in this comparison, we only consider the 
third scenario structures as the others are not candidates for 
being groundstates, according to the additive model. Comparing 
the two version of the MJ potentials, we are also able to look at 
the same groundstates belonging to the second scenario. We found 
a similarity of about 36\% in this case which is less than the 
same study of the third scenario. This conclusion is consistent 
with our previous observation that these structures possess 
smaller energy gaps. This is another fact that lets us consider 
structures with non-degenerate contact vectors as reasonable 
candidates for the groundstates of protein-like structures. 
 
Although the additive potential is not successful in evaluating 
the same groundstates found by MJ potential, we may still 
investigate its capability in finding the general properties of 
the structures such as the spectrum of designability. The 
designability is defined as the number of sequences folding into 
a specific structure (see for example~\cite{22}). 
Figure~\ref{nscompare} is a comparison of the designabilities of 
compact structures evaluated by the additive model 
($n_{s}^{\prime}$) with the evaluations of the MJ model ($n_{s}$) 
for $L=36$. As mentioned before, we have ignored the structures 
with $g>1$. Concluding from the corresponding figure, the 
parameters mentioned above strongly correlate with each other 
($r=0.956$). The correlation coefficient for compact structures 
with $L=16,20,30$ are $0.996, 0.992, 0.959$, respectively. 
Considering the whole structure space for $L=14,16$, the 
designabilities still show strong correlations of $0.997$ and 
$0.996$ respectively. Designabilities of structures in two 
different versions of the MJ potential (MJ96 and MJ85) also agree 
with each other($r=0.994$, Figure~\ref{nscompare}), meaning that 
the relative designability of structures is considerably 
insensitive to small perturbations in interactions. 
 
Additionally, it is also notable that the first few highly 
designable structures are similar in both models, which means 
that the additive potential is able to indicate highly designable 
structures successfully. It should be noted that yielding to 
these structures is considerably more expensive using the 
reference pair-contact potential. The conservation of 
designability order has been previously reported in the case of 
two-letters model~\cite{4}. Looking at the difference between the 
designabilities of structures in the additive and reference 
potentials, (Figure~\ref{hisns-nsprime}) we notice that it is 
highly peeked near zero, implying that it is possible to indicate 
the designability of structures using the additive model with a 
small relative error in most cases. However the tail of the 
histogram in Figure~\ref{hisns-nsprime} warns for the possibility 
of evaluating substantially different designabilities. A closer 
look at Figure~\ref{nscompare} ensures us that the order (rank) of 
designability is more conserved, specially for highly designable 
structures. Therefore, one can conclude that the additive 
potential is more successful in the study of the statistical 
characteristics of the structures than single designs. 
 
\subsection{A crossing point in designability-ranking graphs} 
 
We have also observed that regardless of the used potential, 
designability obeys a universal law in terms of ranking. 
Figure~\ref{ranking} reflects the designability of all structures 
as a function of their rank for different potentials, including 
three versions of the MJ interactions introduced by Miyazawa and 
Jernigan in 1985, 1996 and 1999 and also the additive potential 
extracted from MJ96. In all of these models, the graphs are 
linear in a semi-log plot (Figure~\ref{ranking}-a), specially for 
highly designable structures. Assuming that 
\begin{equation} 
\label{Zipf} r=A e^{-\lambda n_{s}}, 
\end{equation} 
where $r$ and $n_{s}$ are the rank of a specific structure and 
its designability respectively and $\lambda$ is a model-dependent 
free parameter. 
 
The normalization factor $A$ may be determined by noting that the 
summation over designability of all structures equals to the 
number of studied sequences, $S$. That is, 
\begin{eqnarray} 
S = \sum_{r=1}^{N} n_{s}(r) & =&\frac{1}{\lambda}\sum_{r=1}^{N} (\ln A- \ln r ), \nonumber \\ 
& = & \frac{1}{\lambda} (N \ln A - \ln (N!)  ), 
\end{eqnarray} 
where $N$ is the number of structures. Solving the last equation 
for $A$, employing the Stirling's approximation and substituting 
it into Eq.~(\ref{Zipf}), we finally reach to: 
\begin{equation} 
\label{rankdes} r= N e ^{\frac{\lambda S}{N}-1-\lambda n_{s} }= 
\frac{N}{e} e ^{-\lambda (n_{s}-\overline{n}_{s}) }. 
\end{equation} 
Where $\overline{n}_{s} = S/N$ denotes the average designability 
of the structures. 
 
Equation~(\ref{rankdes}) introduces a fix point on 
$n_{s}=\overline{n}_{s}$, meaning that the rank of structure with 
$n_{s}=\frac{S}{N}$ is independent on the model and is equal to 
$\frac{N}{e}$ (see Figure~\ref{ranking}). By rescaling the 
rankings with respect to the total number of structures (which is 
different for additive and non-additive potentials) and 
subtracting $\overline{n}_{s}$ from the designabilities, the 
fixed point appears more visible (Figure~\ref{ranking}-b). The 
same feature has been previously observed for two-, four- and 
infinite-letters models~\cite{91}. 
 
\subsection{The first excited state energy gap} 
 
It is believed that the native states of most proteins are stable 
against thermal fluctuations and mutations. This stability is a 
result of large gap between the free energy of the native states 
and the first excited states, which is a characteristic of the 
protein-like sequences. If we approximate the difference of free 
energy with the energy gap of states, we expect a positive 
correlation between the averages of the energy gaps of the 
sequences which fold into a specific structure, 
$\langle\Delta\rangle$, and the designability of the 
corresponding structure. 
 
Comparing the average energy gap of different structures in the 
additive and the reference potential yields to quite similar 
diagrams (Figure~\ref{gapns}), although the energy gaps in 
additive model are slightly smaller than those of MJ model. The 
difference of the number of points of the additive and MJ graphs 
is a result of ignoring structures having $g>1$ in additive 
model. There is a jump in the energy gap diagram  of short 
sequences which may be used to distinguish low and high 
designable structures from each other (e.g. length 16 in 
Figure~\ref{gapns}-a and b). This transition becomes smoother for 
sequences with larger length (e.g. length 36 in 
Figure~\ref{gapns}-c and d). The jump in energy gap is a 
distinguished characteristic of finite size structures which is 
more visible in the models with less free parameters (such as the 
HP model) than the complicated models (such as MJ reference 
potential)~\cite{22,92}. However, the highly designable 
structures possess a larger energy gap in average, regardless of 
the used model. 
 
Figure~\ref{probdel} reports the probability of finding the same 
structure as groundstate in different models as a function of 
energy gap, $\Delta$. The probability increases monotonically as a 
function of $\Delta$ which means that deeper groundstates of the 
reference potential have a higher chance to be groundstates in 
the additive model too. 
 
Marking the groundstate and the first excited structure of each 
sequence in the MJ potential model and comparing the folding 
energy difference in this model, $\Delta$, with the energy 
difference in the additive model, $\Delta^{\prime}$, we reach to 
a meaningful correlation between them. If these structures had 
been chosen randomly, we would have expected a Gaussian 
distribution for $\Delta - \Delta^{\prime}$ centered in origin. 
However, this difference has a non-zero mean centered at $0.11 
k_{\rm B}T$ (Figure~\ref{hisdeldelprime}) for these models. The 
small tendency of these structures to fold in the same 
groundstate is associated to their smaller energy differences in 
the additive model. 
 
\subsection{Protein design using additive potential} 
 
Designing a protein can be described as finding a sequence, 
$\sigma$, which maximizes the Boltzmann probability of finding 
the protein in a specific configuration, $S^*$. Therefore, one 
must maximize: 
\begin{equation} 
P= \frac{e^{\frac{-H(\sigma,S^{*})}{k_{\rm B}T}}}{\sum 
e^{\frac{-H(\sigma,S^{i})}{k_{\rm B}T}}} \ \ , 
\end{equation} 
with respect to $\sigma$, where $H(\sigma,S^{*})$ is the folding 
energy of $\sigma$ in structure $S$. The additive potential may 
be used as a first approximation in protein design. We have 
studied the ability of the additive model to design sequences for 
all 69 maximally compact $2D$ square lattice structures with 
$L=16$. This attempt failed for $24$ of them, basically due to 
the degeneracy of their contact vectors ($g>1$), together with 
$6$ other structures with $g=1$. In these cases, the sequence 
which maximizes the probability has degenerate groundstate. For 
the remaining $39$ structures, we could successfully design a 
sequence to fold in the additive potential model. Moreover, the 
evaluated groundstate structures remains unchanged under the 
application of MJ potential. The energy gap of these sequences 
varies in the range $2.90k_{\rm B}T$ to $8.56k_{\rm B}T$ and the 
mean value of the gaps is equal to $5.91k_{\rm B}T$. 
 
\section{Conclusion} 
 
We have shown that the additive part of the pair-contact 
interactions between amino acids (introduced by Miyazawa and 
Jernigan~\cite{52}) essentially represents the main 
characteristics of the reference interactions. Taking the 
additive part as the leading term, we have shown that the 
deviations may be considered as a first-order perturbation to the 
reference interactions. The additive (solvation) model introduces 
less energy parameters while reduces the encoding of structures 
from a matrix form to a vector form. 
 
Considering the MJ96 potential as the reference potential, we 
have investigated the conservation of the groundstates of the 
structures under employing the additive model. The groundstates of 
about one fourth of sequences are conserved in the case of 
$6\times6$ structures. This number is improved to one third if we 
consider only non-degenerate ($g=1$) structures. 
 
The additive potential is quite successful in predicting the 
statistical properties of the structures. The designability of the 
structures evaluated by the reduced model is highly correlated to 
the reference model. The additive model successfully indicates the 
highly designable structures (evaluated by the reference 
potential), all of them being among $g=1$ structures. This 
observation suggests that non-degenerate vectors (structures) 
extensively resemble the protein-like structures. Sequence design 
for such structures is also successful using the additive model. 
 
Taking the other version of MJ matrix (MJ85) as another 
perturbation to the MJ96 results in a same magnitude of 
deviation, although the computation is much cheaper using the 
additive model. 
 
We have also observed a crossing point in the log-linear diagram 
of designability ranking, showing that about $\frac{1}{e}$ of the 
structures have designabilities above the average, independent on 
the used model. 
 
\begin{acknowledgments} 
We would like to thank M. Babadi and B. Mehmani for carefully 
reading the manuscript and useful comments. 
\end{acknowledgments} 
 
\bibliography{bibtex} 
 
\begin{table*} 
\caption{\label{reg} Correlation between different hydrophobicity 
scales~\cite{hydro} (The upper half represent the correlation 
coefficients and the lower half is the p-value multiplied by 
$10^{5}$)} 
\begin{ruledtabular} 
\begin{tabular}{llcccccccccccr} 
Hyd. scale & ADD96 & ADD85 & HS1 & HS2 & HS3 & HS4 & HS5 & HS6 & HS7 & HS8 & Li \textit{et al}\\ 
\colrule 
ADD96 & - & 0.979 & 0.790 & 0.830 & 0.758 & 0.818 & 0.724 & 0.759 & 0.836 & 0.914 & 0.996\\ 
ADD85 & 1 & - & 0.783 & 0.806 & 0.758 & 0.807 & 0.744 & 0.739 & 0.834 & 0.918 & 0.979 \\ 
HS1 & 41 & 46 & - & 0.879 & 0.686 & 0.760 & 0.668 & 0.724 & 0.776 & 0.875 & 0.808\\ 
HS2 & 21 & 31 & 9 & - & 0.587 & 0.723 & 0.452 & 0.751 & 0.705 & 0.810 & 0.857\\ 
HS3 & 70 & 70 & 214 & 864 & - & 0.925 & 0.787 & 0.721 & 0.718 & 0.835 & 0.759\\ 
HS4 & 25 & 31 & 67 & 123 & 4 & - & 0.655 & 0.930 & 0.801 & 0.887 & 0.828\\ 
HS5 & 121 & 88 & 281 & 4324 & 43 & 338 & - & 0.434 & 0.754 & 0.780 & 0.698\\ 
HS6 & 68 & 95 & 121 & 78 & 127 & 3 & 5230 & - & 0.767 & 0.811 & 0.776\\ 
HS7 & 19 & 19 & 52 & 161 & 132 & 34 & 75 & 60 & - & 0.870 & 0.826\\ 
HS8 & 4 & 4 & 9 & 29 & 19 & 7 & 49 & 29 & 10 & - & 0.920\\ 
Li \textit{et al} & 1 & 1 & 30 & 13 & 68 & 21 & 181 & 52 & 22 & 4 & - \\ 
\end{tabular} 
\end{ruledtabular} 
\end{table*} 
 
\begin{table*} 
\caption{\label{Lattice} The number of contact matrices and 
contact vectors of maximally compact structures on 2D square 
lattices} 
\begin{ruledtabular} 
\begin{tabular}{llcccccccccccr} 
Length & 16 & 20 & 30 & 36 \\ 
\colrule 
Structures & 69 & 503 & 13498 & 57337\\ 
Vectors & 56 & 398 & 9514 & 35662\\ 
Vectors with g=1 & 45 & 309 & 6819 & 23921\\ 
Average of g & 1.23 & 1.26 & 1.42 & 1.61\\ 
\end{tabular} 
\end{ruledtabular} 
\end{table*} 
 
\begin{table*} 
\caption{\label{scenarios} Probability of occurrence of each 
scenario with their average energy gap } 
\begin{ruledtabular} 
\begin{tabular}{llcccccccccccr} 
Length & 16 & 20 & 30 & 36\\ 
\colrule 
Probability of first scenario & 3.8\% & 3.2\% & 3.1\% & 3.3\% \\ 
Probability of second scenario & 23.4\% & 24.7\% & 33.1\% & 37.4\% \\ 
Probability of third scenario & 72.8\% & 72.1\% & 63.8\% & 59.3\% \\ 
Average energy gap (second scenario) [$k_{\rm B}T$] & 0.34 & 0.32 & 0.33 & 0.34 \\ 
Average energy gap (third scenario) [$k_{\rm B}T$] & 0.92 & 0.69 & 0.57 & 0.54 \\ 
\end{tabular} 
\end{ruledtabular} 
\end{table*} 
 
\begin{table*} 
\caption{\label{probability2} The ratio of invariant groundstates 
of the reference potential under the application of variant 
potentials} 
\begin{ruledtabular} 
\begin{tabular}{llcccccccccccr} 
&Additive96 &MJ85 &Additive85    \\ 
\colrule Second Scenario & 10.5\% & 35.8\% & 9.0\% \\ Third 
Scenario & 33.0\% & 43.4\% & 28.5\% \\  Total \footnote{The total 
probability is not simply the sum of above numbers but their 
average with respect to the weights given in table 
\ref{scenarios}.}  & 24.5\% & 40.4\% & 
21.0\% \\ 
\end{tabular} 
\end{ruledtabular} 
\end{table*} 
 
\newpage 
 
{\bf Figure captions:} 
 
\begin{figure} 
\includegraphics[width=15cm]{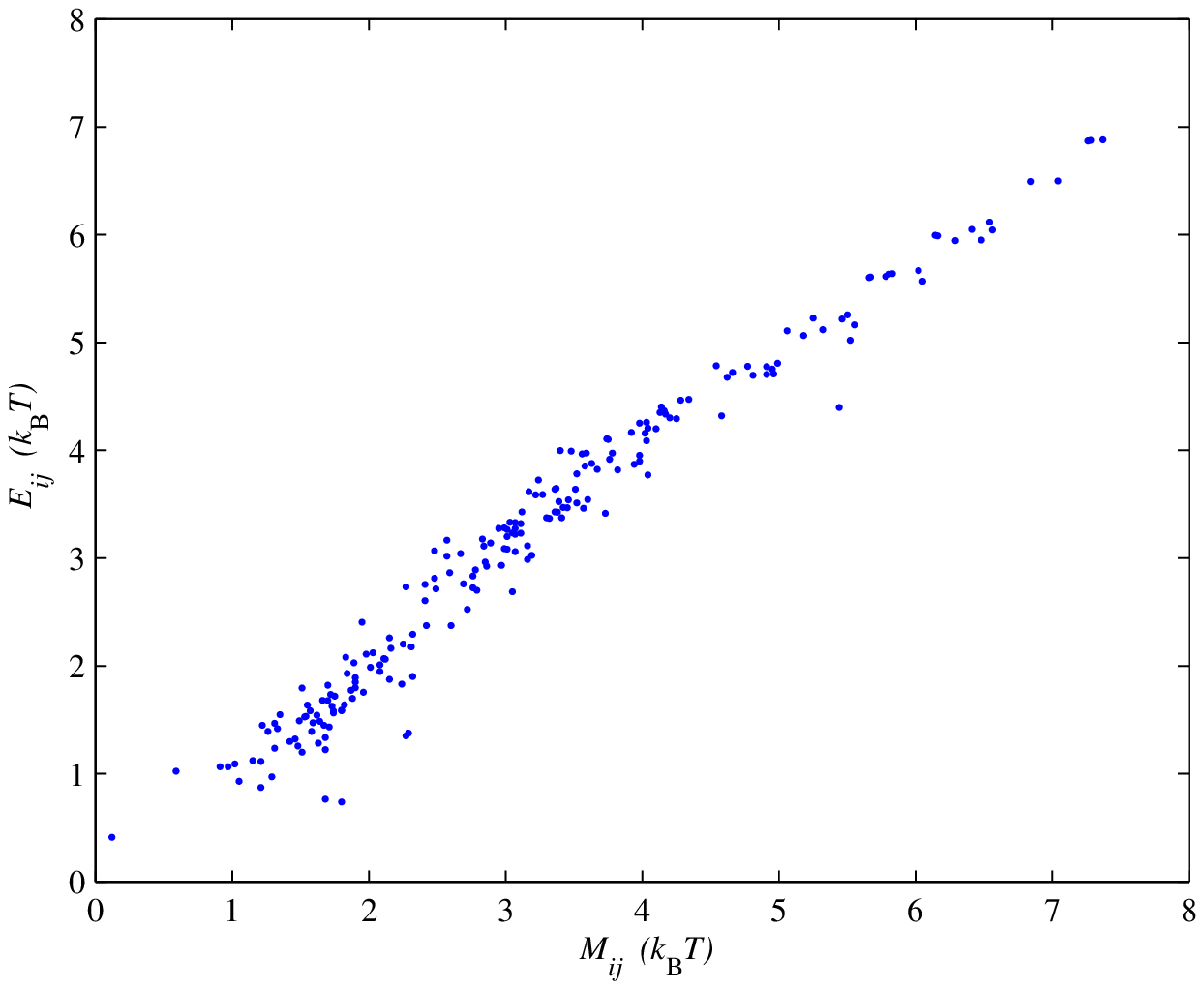} 
\caption {\label{mjaddmat} } 
\end{figure}

\begin{figure} 
\includegraphics[width=15cm]{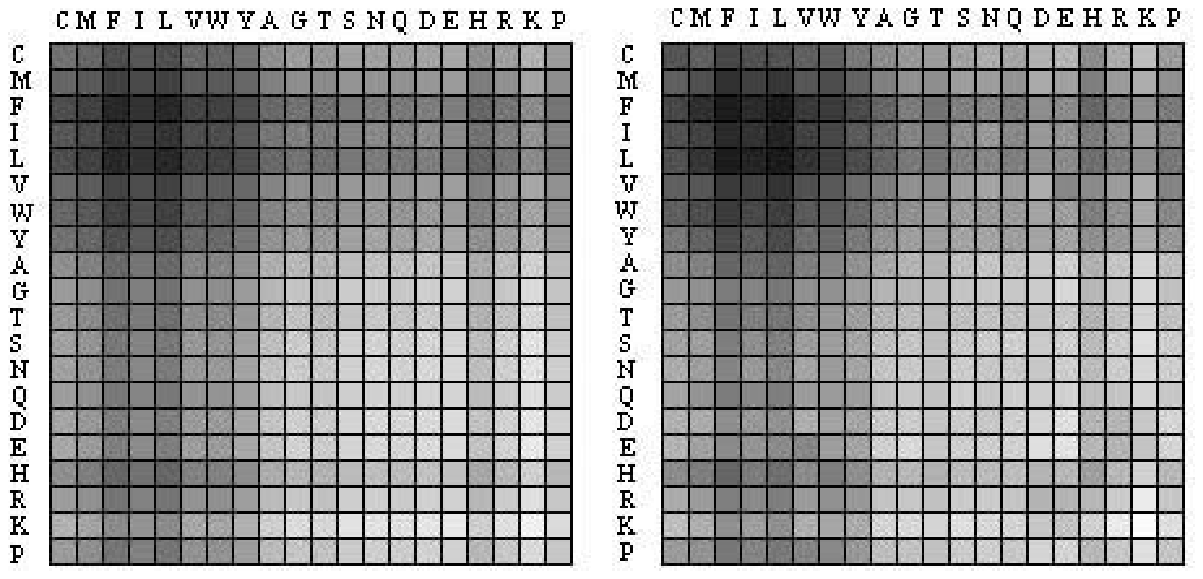} 
\caption {\label{gray} } 
\end{figure}

\begin{figure} 
\includegraphics[width=15cm]{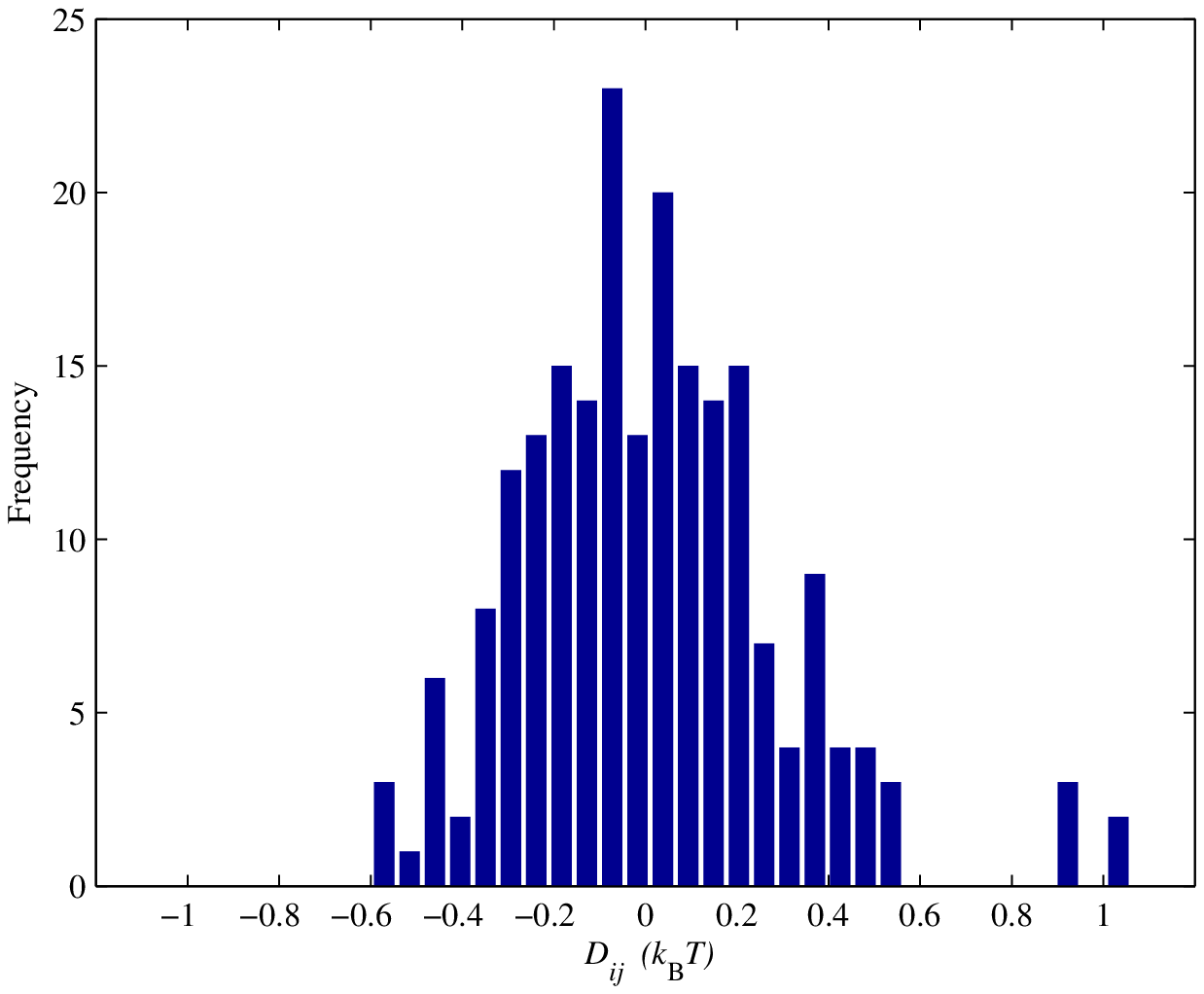} 
\caption {\label{diff} } 
\end{figure}

\begin{figure} 
\includegraphics[width=15cm]{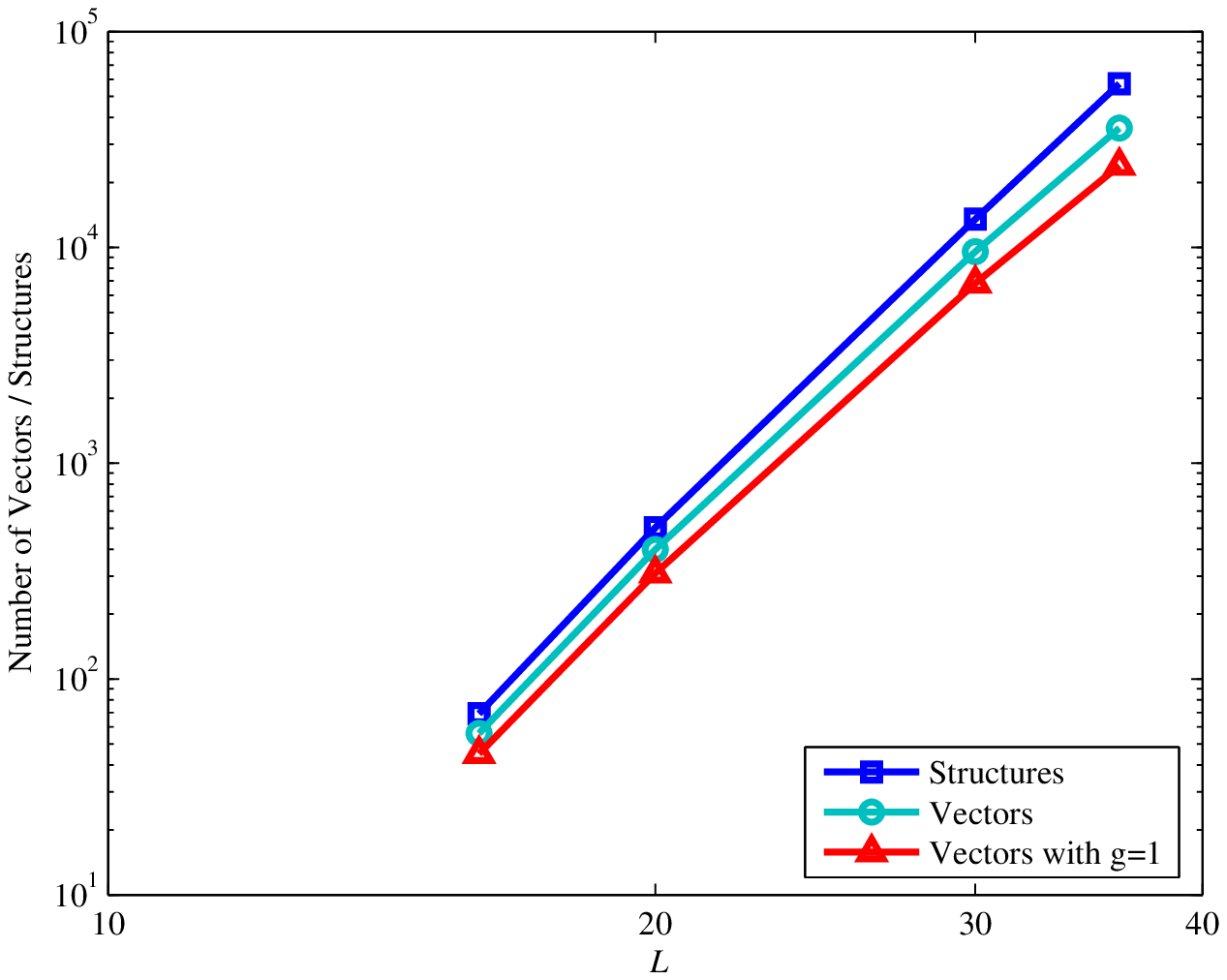} 
\caption {\label{lenghts} } 
\end{figure}

\begin{figure} 
\includegraphics[width=15cm]{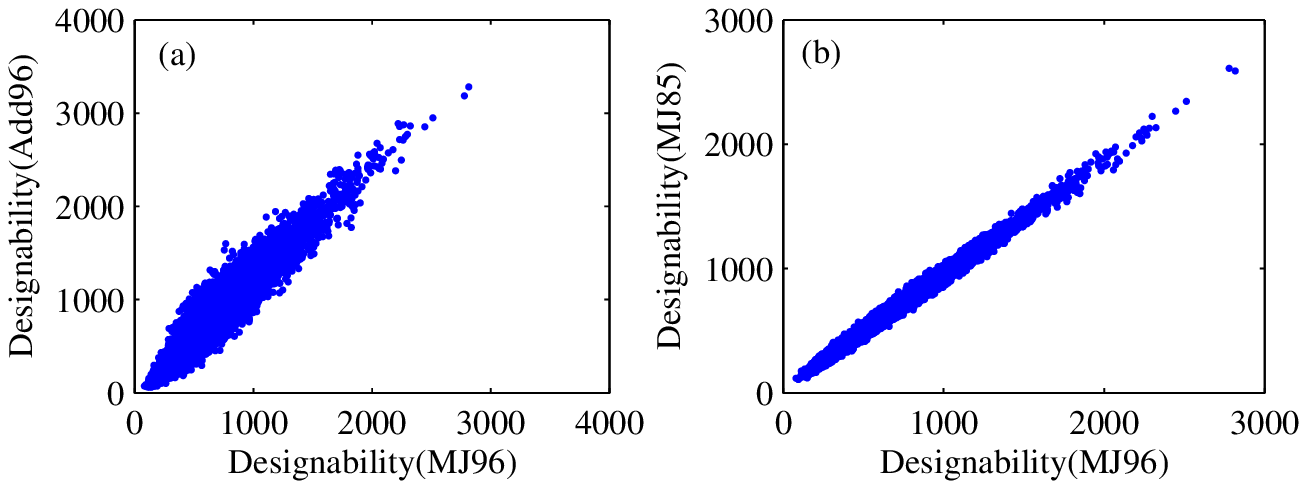} 
\caption {\label{nscompare}} 
\end{figure}

\begin{figure} 
\includegraphics[width=15cm]{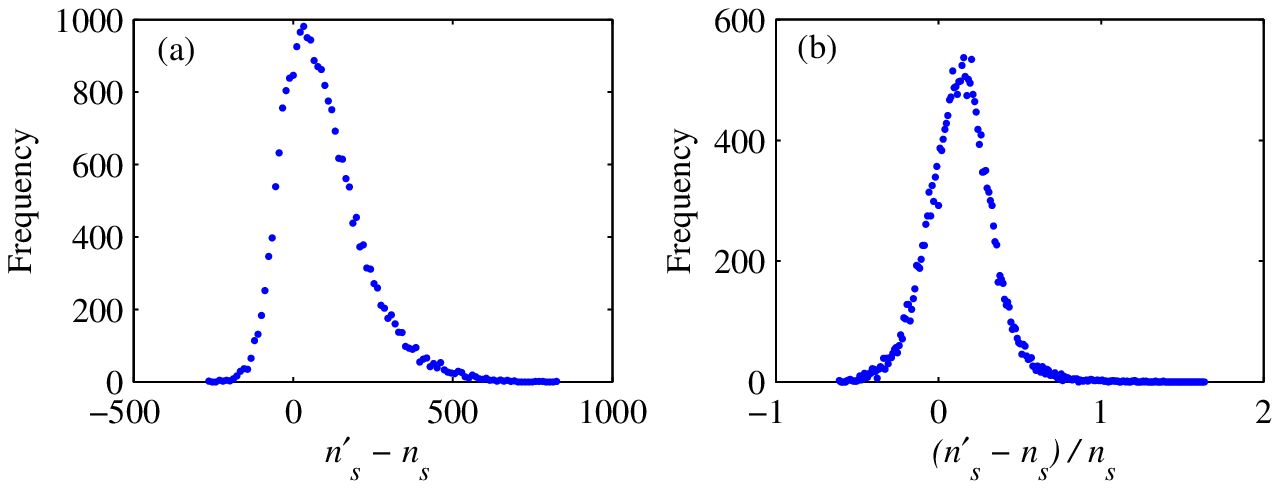} 
\caption {\label{hisns-nsprime}} 
\end{figure}

\begin{figure} 
\includegraphics[width=15cm]{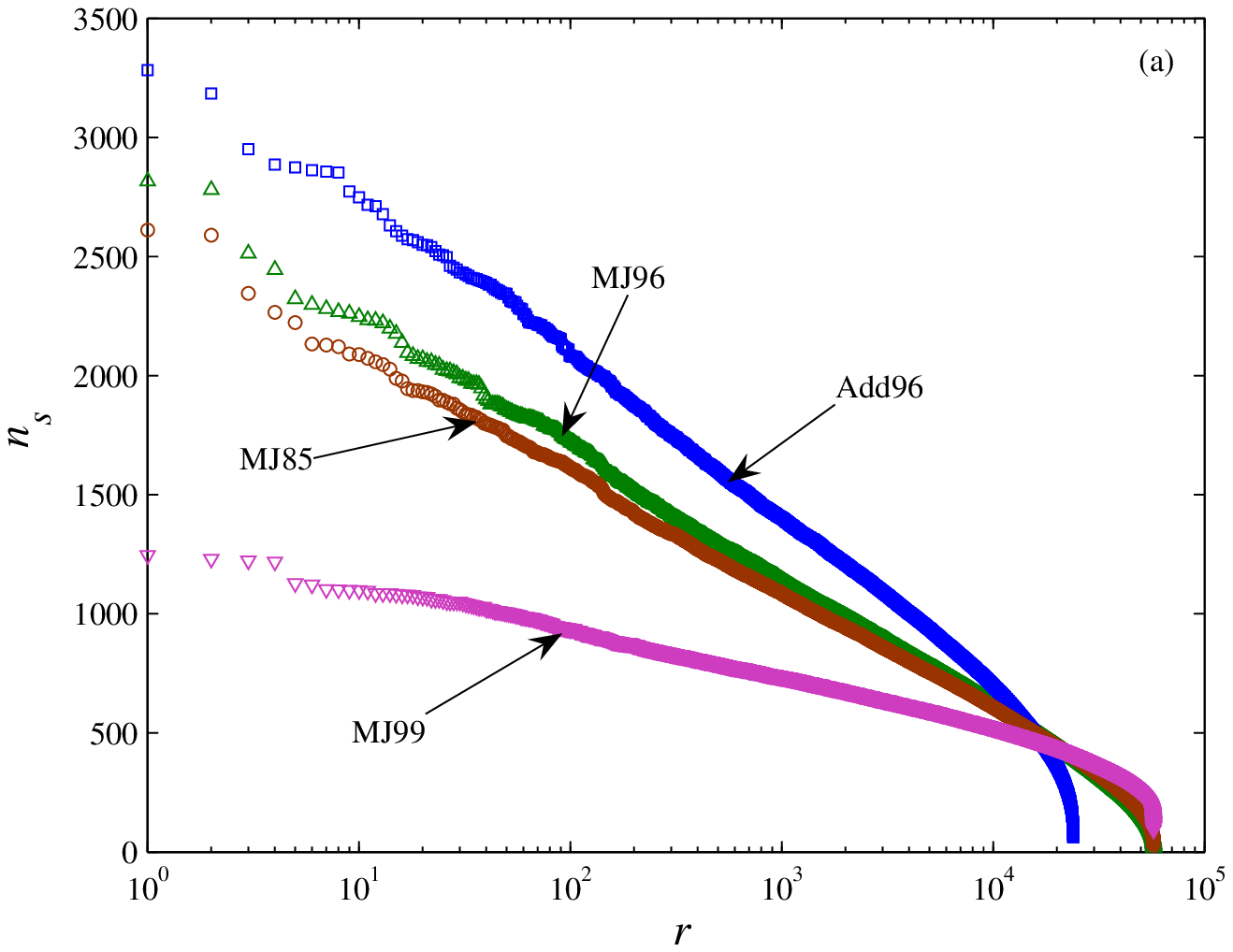} 
\includegraphics[width=15cm]{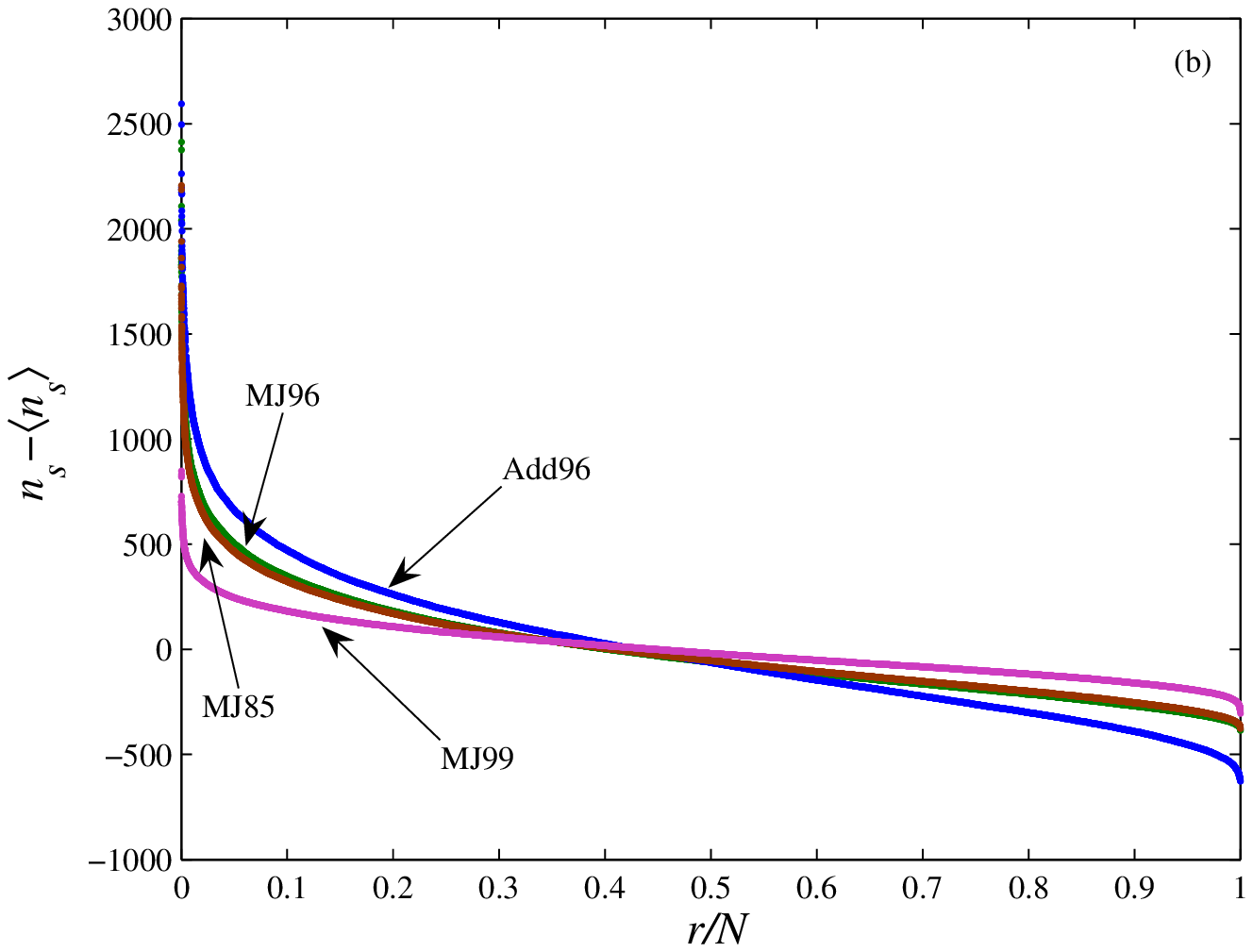} 
\caption {\label{ranking}} 
\end{figure}

\begin{figure} 
\includegraphics[width=15cm]{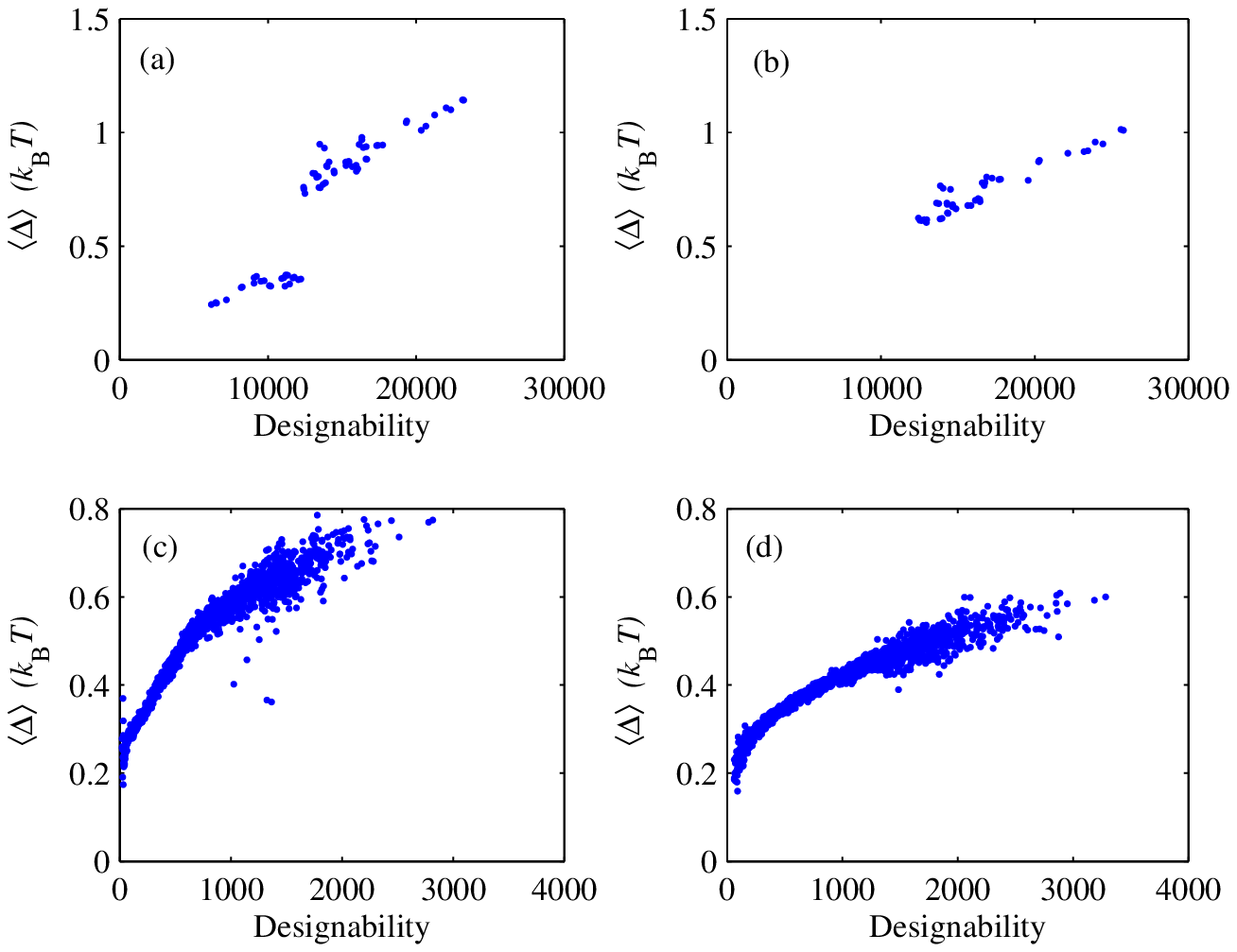} 
\caption {\label{gapns} } 
\end{figure}

\begin{figure} 
\includegraphics[width=15cm]{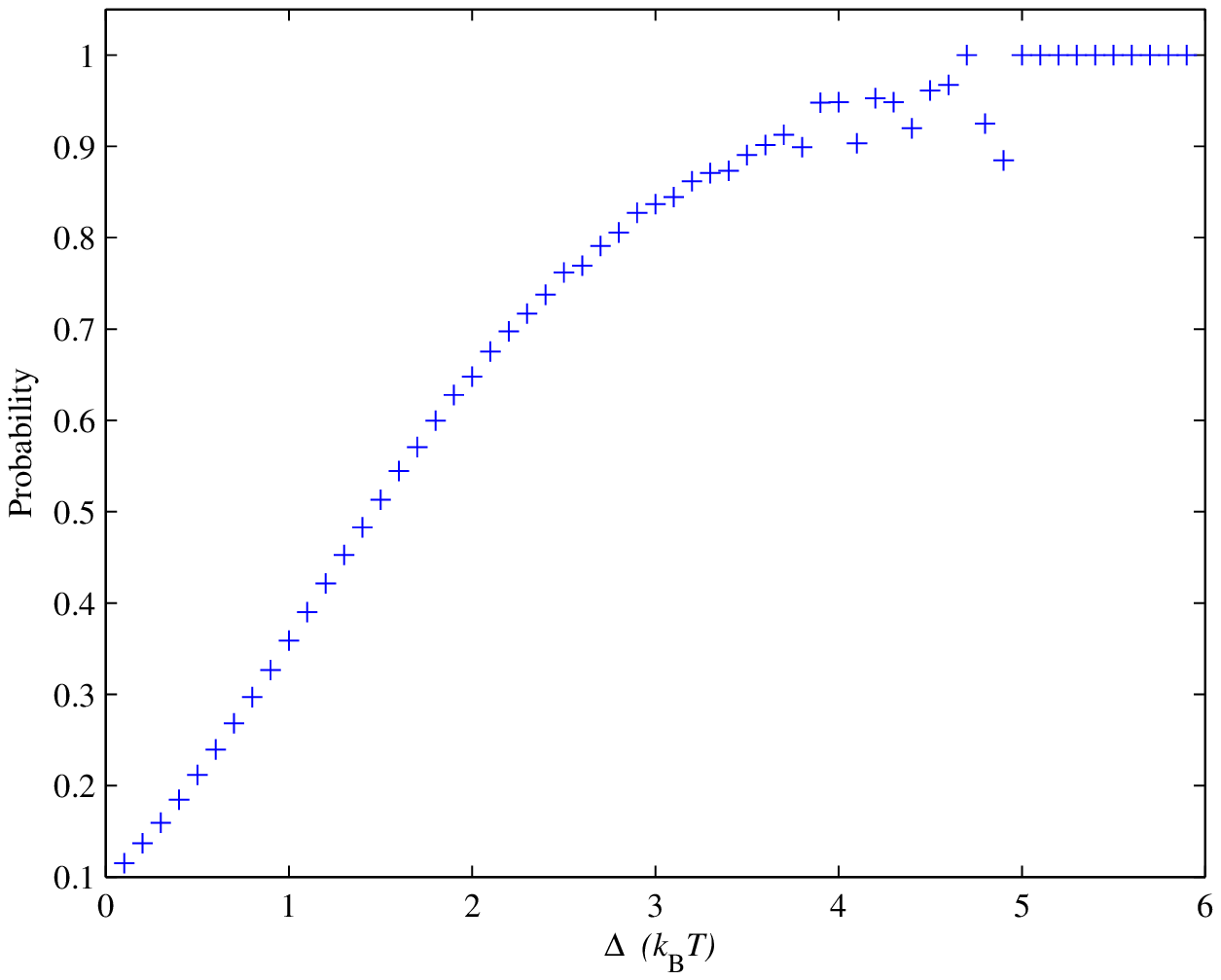} 
\caption {\label{probdel} } 
\end{figure}

\begin{figure} 
\includegraphics[width=15cm]{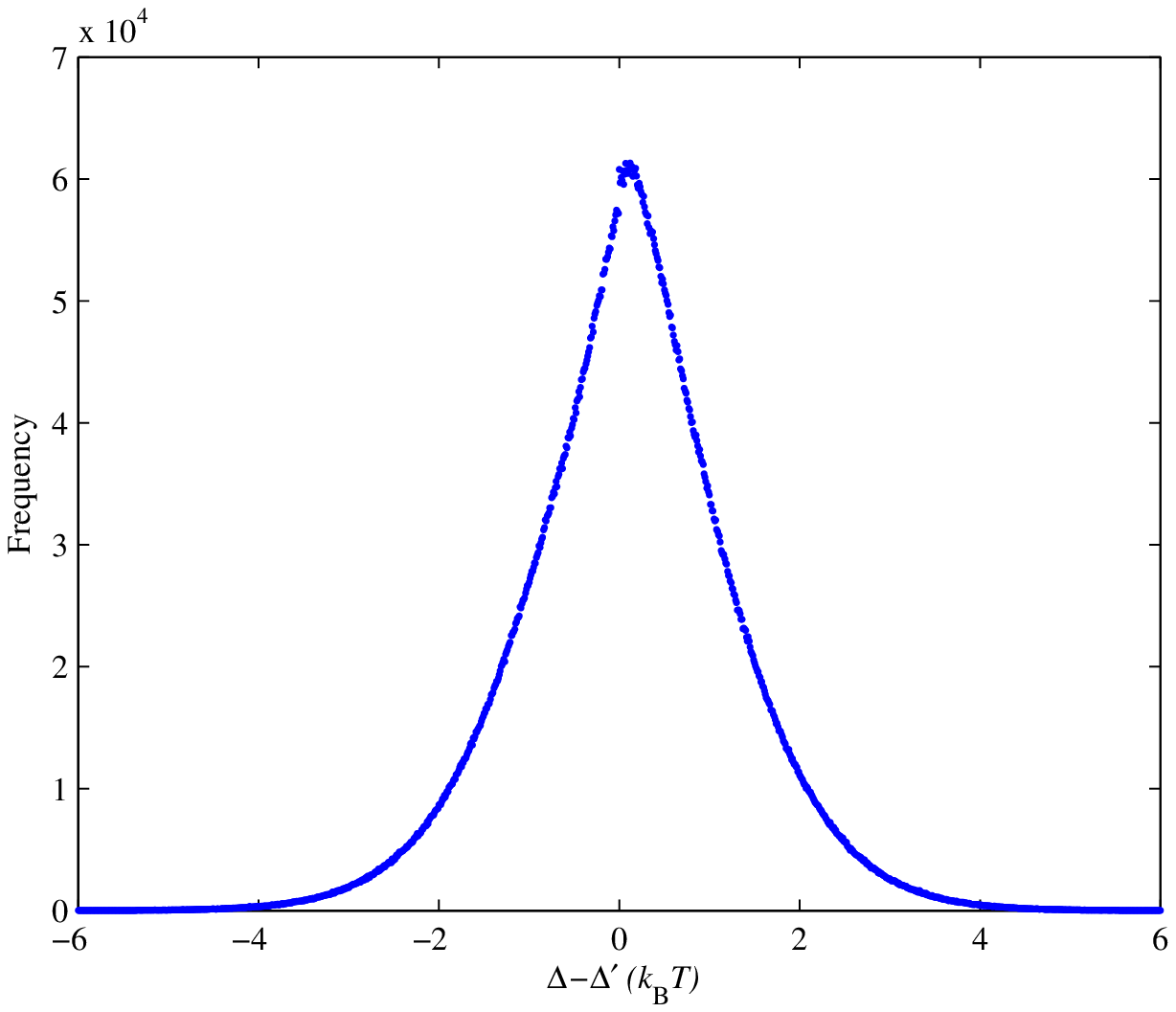} 
\caption {\label{hisdeldelprime} } 
\end{figure}

\begin{itemize} 
 
\item Figure~[\ref{mjaddmat}]: The reference potential (MJ96) and 
the extracted additive potential (Additive96) parameters are 
compared with each other, showing a strong correlation 
($r=0.982$). 
 
\item Figure~[\ref{gray}]: Graphical comparison of the reference 
potential (left) and extracted additive potential (right). Darker 
elements correspond to stronger interactions. 
 
\item Figure~[\ref{diff}]: 
The histogram of the elements of $D$ (the difference of 
Additive96 and MJ96 matrices).  Except a few elements, the rest 
of the differences are less then $0.5k_{\rm B}T$. 
 
\item Figure~[\ref{lenghts}]: Color online: The number of highly compact 
structures, their corresponding contact vectors and the number of 
vectors with $g=1$ for different lengths are presented in a 
log-log plot. 
 
\item Figure~[\ref{nscompare}]: The designability of a given 
structure in the reference model (MJ96) is compared with its 
designability in two other models: (a) Additive96 (b) MJ85. The 
most designable structures are the same in both cases. 
 
\item Figure~[\ref{hisns-nsprime}]: (a) The histogram of difference 
of designabilities of a given structure evaluated by 
Additive96,$n_{s}^{\prime}$, and MJ96, $n_{s}$. (b) The histogram 
of the relative change in the designabilities. The difference in 
designabilities can be neglected, compared to the designabilities 
themselves. 
 
\item Figure~[\ref{ranking}]: Color online: The structures are sorted in 
descending order of their designabilities in different models. 
(a) Designability almost is a log-linear function of rank in all 
studied models. (b) The same plot in a linear-linear scale. The 
vertical axis is shifted by mean of designability of the 
structures. The horizontal axis also scaled by the total number 
of structures. All of the graphs pass through a fixed point. 
 
\item Figure~[\ref{gapns}]: The average energy gap between the 
groundstate and the first excited state of a given structure is 
compared to its designability in different models. (a) $4\times4$ 
structures in the MJ96 potential. (b) $4\times4$ structures in 
the Additive96 potential. (c) $6\times6$ structures in the MJ96 
potential. (d) $6\times6$ structures in the Additive96 potential. 
In the case of additive models (b and d), the structures with 
$g>1$ (degenerate groundstate, $\Delta=0$) have been ignored. A 
jump is observed in subfigures (a) and (b) which is associated to 
the finite size effect (see text). 
 
\item Figure~[\ref{probdel}]: The probability of finding the 
same groundstate for a given sequence using both Additve96 and 
MJ96 models {\it vs.} the energy gap in MJ96 model ($\Delta$). 
 
\item Figure~[\ref{hisdeldelprime}]: The histogram of the change 
in the energy gap of sequences in the reference ($\Delta$ ) and 
the additive ($\Delta^{\prime}$) models. 
 
\end{itemize} 
 
\newpage 
 
\end{document}